\pdfoutput=1

\documentclass[11pt]{article}

\usepackage[final]{acl}
\usepackage{graphicx} 
\usepackage{subcaption} 

\usepackage{times}
\usepackage{latexsym}
\usepackage{amsmath}
\usepackage[T1]{fontenc}

\usepackage[utf8]{inputenc}

\usepackage{microtype}

\usepackage{inconsolata}

\usepackage{graphicx}

\title{DNA Language Model and Interpretable Graph Neural Network Identify Genes and Pathways Involved in Rare Diseases}

\author{Ali Saadat, Jacques Fellay \\
         School of Life Sciences\\
Ecole Polytechnique Fédérale de Lausanne\\
Lausanne, Switzerland\\
\texttt{\{ali.saadat, jacques.fellay\}@epfl.ch}}

\begin{document}
\maketitle
\begin{abstract}
Identification of causal genes and pathways is a critical step for understanding the genetic underpinnings of rare diseases. We propose novel approaches to gene prioritization and pathway identification using DNA language model, graph neural networks, and genetic algorithm. Using HyenaDNA, a long-range genomic foundation model, we generated dynamic gene embeddings that reflect changes caused by deleterious variants. These gene embeddings were then utilized to identify candidate genes and pathways. We validated our method on a cohort of rare disease patients with partially known genetic diagnosis, demonstrating the re-identification of known causal genes and pathways and the detection of novel candidates. These findings have implications for the prevention and treatment of rare diseases by enabling targeted identification of new drug targets and therapeutic pathways.
\end{abstract}

\section{Introduction}

The landscape of genomics research has undergone a profound transformation with the advent of high-throughput sequencing technologies \citep{Metzker2009}. The generation of a vast amount of genomics data offers unprecedented insights into human genetic diversity \citep{Auton2015, Chen2023}. However, this wealth of data brings significant challenges in terms of data analysis and interpretation. A main challenge in deciphering the underlying mechanisms of diseases is establishing a link between genotype and phenotype \citep{Gallagher2018}. This task becomes even harder in the context of rare diseases, where the scarcity of data reduces statistical power \citep{Seaby2020}.

Traditional methods for finding disease-associated genes/pathways have predominantly relied on statistical approaches, such as correlating specific genetic variants with disease occurrence \citep{Auer2015, Uffelmann2021}. These approaches show decent performance if the cohort size is large, which is often a big obstacle in rare disease studies. Moreover, these methods usually utilize basic variant statistics (such as number of variant carriers), and might not take into account the gene-specific impact of variants on the gene sequence \citep{MacArthur2014}. 

Another family of computational approaches for gene/pathway prioritization rely on the concept of guilt-by-association, where genes/pathways are considered potentially relevant based on their similarity to known disease genes \citep{Lee2011, Guala2017}. These methods work well in scenarios where some underlying genetic factors of the phenotype are well-studied, which is not the case for many diseases \citep{Amberger2018}. Moreover, these methods might introduce bias since they look for similar genes, thereby missing novel disease-causing genes \citep{Gillis2012}.

Recent years have seen a remarkable rise in the performance of language models, particularly in the field of natural language processing (NLP) \citep{Delvin2018, radford2019language}. These models ‘learn' language by processing vast amounts of text data, enabling them to perform a wide range of downstream tasks such as translation, summarization, and question-answering with unprecedented accuracy and fluency \citep{Zhao2023}. Parallel to this development, the concept of language models has been applied to genomics, giving rise to DNA language models (DNA-LMs) \citep{Zhihan2023, DallaTorre2023, Benegas2023, Nguyen2023}. Genomic sequences, much like textual data, comprise long chains of information, in this case nucleotides instead of words. DNA-LMs apply the principles of NLP to interpret and analyze these sequences, translating the 'language' of DNA into meaningful biological insights. By learning from extensive genomic data, these models can provide new perspectives on downstream biological processes \citep{Consens2023, Marin2023}. 

HyenaDNA \citep{Nguyen2023} is a long-range genomic foundation model pre-trained on the human reference genome at single nucleotide resolution. It can process long-range DNA sequences and represent them as embeddings in a high-dimensional space. For any genomic region such as a gene, HyenaDNA generates embeddings that capture the inherent information of the DNA sequence. These embeddings dynamically change in response to genetic variants, offering insights into how genetic alterations impact biological processes. 

\begin{figure}[t]
    \centering
    \begin{subfigure}[b]{\columnwidth} 
        \includegraphics[width=\columnwidth]{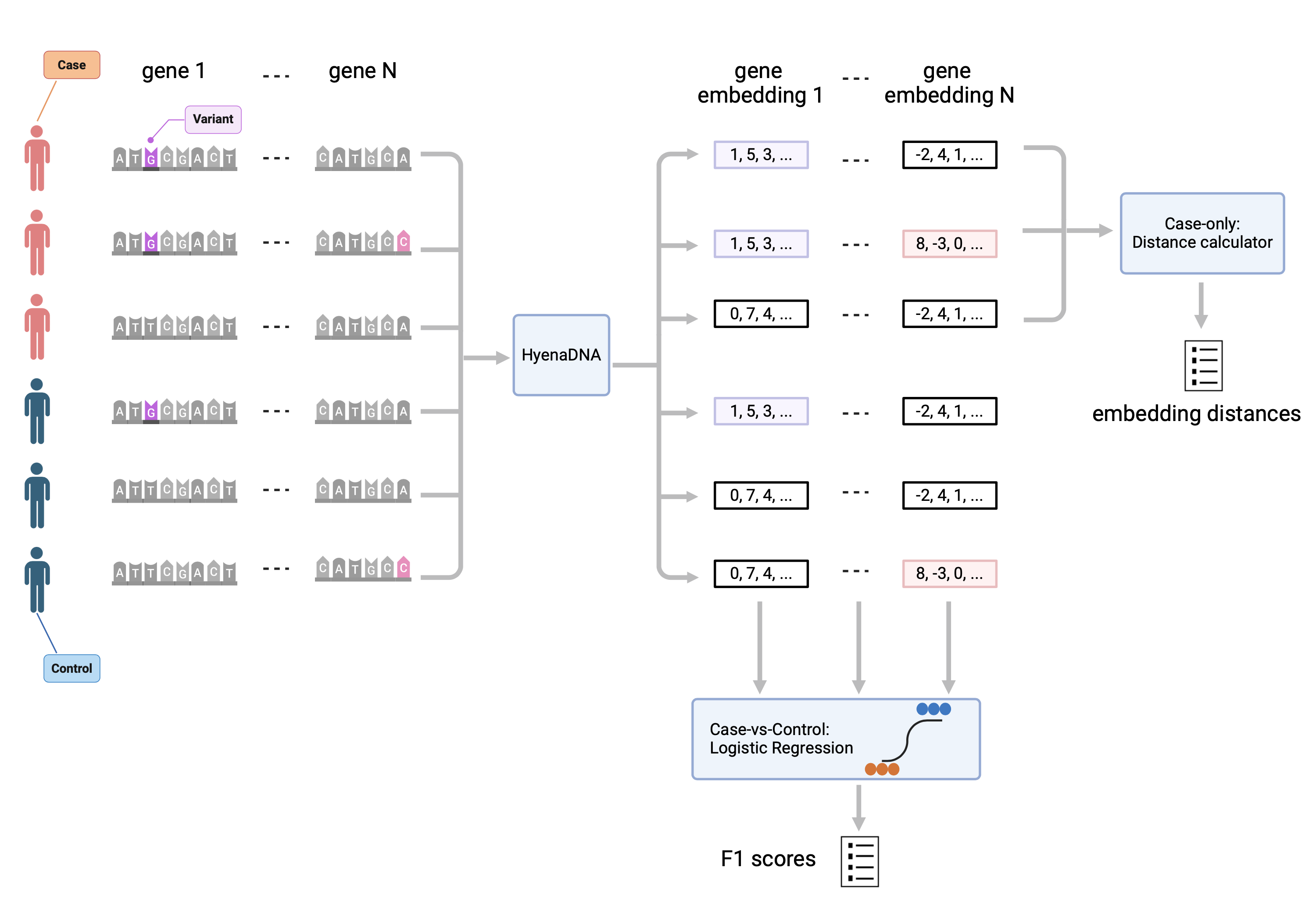} 
        \caption{Gene prioritization workflow: DNA sequences of candidate genes are passed to HyenaDNA for gene embedding generation. The embeddings are used to calculate a gene specific score ($F_1$ score for case-vs-control, distance score for case-only), which is used to rank and select top candidate genes.}
        \label{fig:gene_prioritization_workflow}
    \end{subfigure}
    \vskip 0.3in 
    \begin{subfigure}[b]{\columnwidth} 
        \includegraphics[width=\columnwidth]{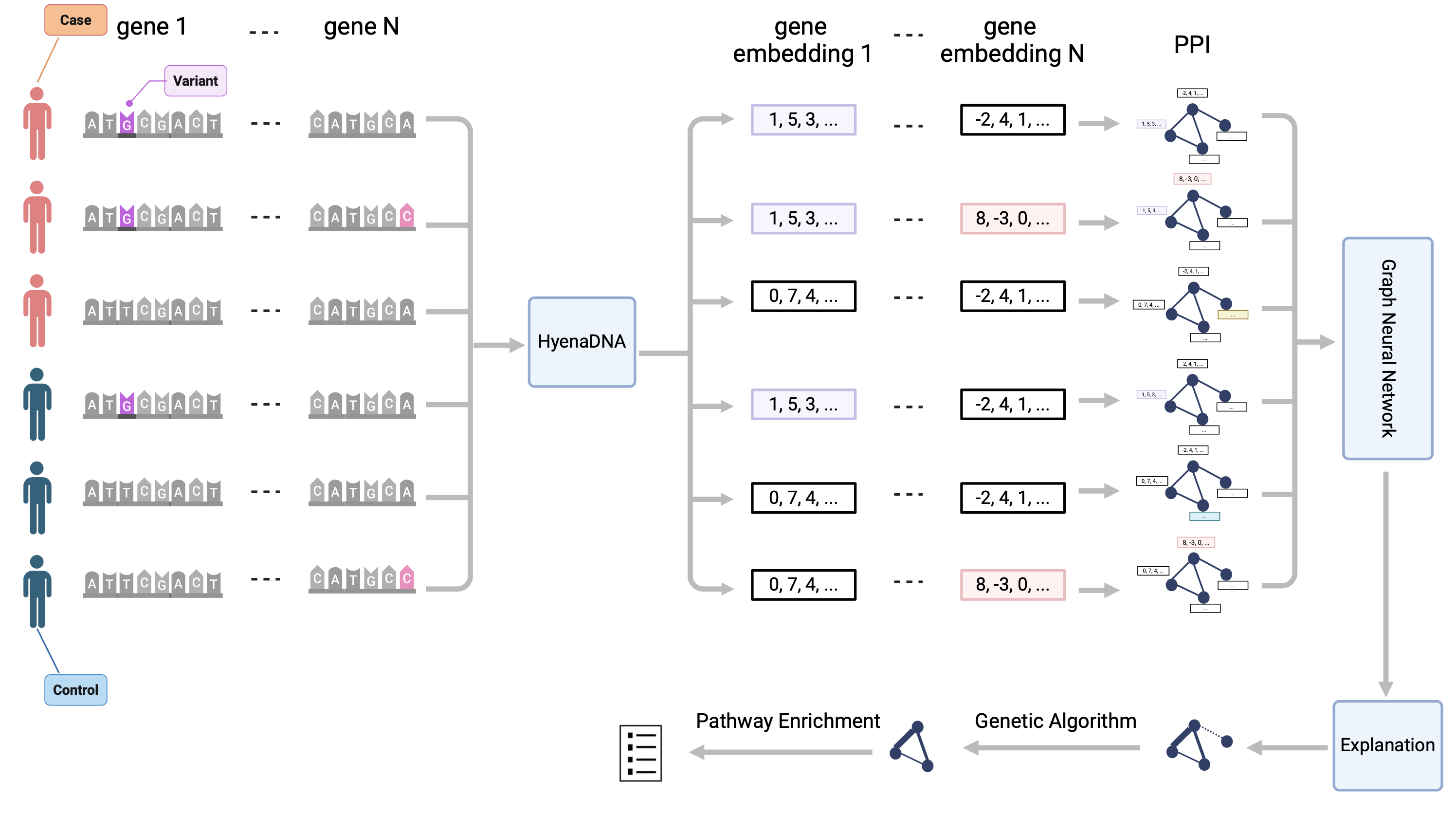} 
        \caption{Pathway identification workflow: for each individual, a protein-protein interaction network is constructed with gene embeddings as node features. A GNN is trained to classify cases graphs from controls, and GNNExplainer is applied to score the importance of nodes and edges for graph classification. Afterwards, a Genetics Algorithm is used to find the most explainable subnetwork, and a pathway enrichment analysis is performed on that subnetowrk to find the over-represented biological pathways.}
        \label{fig:pathway_identification_workflow}
    \end{subfigure}
    \caption{Overall summary of of the methods. Figures created with \url{BioRender.com}.}
    \label{fig:combined}
\end{figure}

We hypothesize that variants with strong deleterious effects have a detectable impact on gene embeddings. We designed complementary methods to identify genes and pathways that contain such deleterious variants and could therefore play a causal role in the pathogenesis of rare diseases. For gene prioritization, we propose two approaches (case-vs-control and case-only) to quantitatively rank candidate genes (Figure \ref{fig:gene_prioritization_workflow}). For pathway identification, we propose a method that combines DNA-LM, interpretable graph neural networks (GNN) \citep{Wu_2021,ying2019gnnexplainer} and Genetic Algorithm \citep{Katoch2020} (Figure \ref{fig:pathway_identification_workflow}).  We validate our methods on a cohort of rare disease patients with partially known genetic diagnosis, demonstrating the re-identification of known causal genes and the detection of novel candidates.

\section{Methods}

\subsection{Study participants}

We selected two cohorts from our in-house database of exome-sequenced individuals. The first cohort consists of 120 previously healthy children who were admitted to pediatric intensive care units (PICUs) with respiratory failure due to a common viral respiratory infection. This cohort serves as the “rare disease” patient group for this study. As control group, we selected a total of 172 healthy individuals. The studies were approved by the relevant ethics commissions and all study participants provided a signed informed consent for research including human genetic testing. 

\subsection{Short-read alignment and variant calling}

Adapter sequences were trimmed from sequencing reads using fastp \citep{Chen2018} and the reads were subsequently aligned against the human reference genome (hg38) using the maximum exact matches algorithm in Burrows-Wheeler Aligner \citep{Li2009}. The Genome Analysis Software Kit (GATK4) best-practice pipeline was used to call variants in the multi-sample mode \citep{DePristo2011}. In summary, PCR duplicates were removed and base quality scores were recalibrated to correct for sequencing artifacts. We called individual-level variants with GATK HaplotypeCaller before combining single-sample callsets for joint genotyping. To exclude low quality variants, we applied variant quality score recalibration and manual filtering (depth $\geq$ 20, genotype quality $\geq$ 20, and 0.2 $\leq$ heterozygous allele balance $\leq$ 0.8).

\subsection{Variant annotation and filtering}

To predict the potential impact of each variant, we used Variant Effect Predictor (VEP) \citep{McLaren2016}. To identify loss-of-function variants, we used Loss-of-Function Transcript Effect Estimator (LOFTEE) as a VEP plugin \citep{Karczewski2020}.

To classify the variant into putative pathogenicity groups, we implemented the ACMG/AMP guidelines \citep{Richards2015} in R (\url{https://www.r-project.org}) (see full description Appendix \ref{sec:appendix}). A probability of pathogenicity (PoP) was assigned to each variant according to the ACMG/AMP Bayesian classification framework \citep{Tavtigian2018}. Variants with PoP $\geq$ 0.9 were considered as damaging. Genes with at least one pathogenic variant were included in the downstream analysis.

\subsection{Gene embedding calculation}

For candidate gene selection, we kept the genes that passed the following criteria: 1) At least one patient carries $\geq$ 1 pathogenic variant in the gene. 2) The length of the gene (including exons, introns, 3'-UTR, and 5'-UTR) is less than 450,000 nucleotides, which is the maximum input size of the medium-size HyenaDNA. 

For each candidate gene, we obtained the reference gene sequence using biomaRt \citep{Kinsella2011}. Then for each study participant, we altered the reference alleles based on the position of the variants in the gene. The resulting DNA sequence was then fed into the medium-size HyenaDNA to get embeddings for each nucleotide. To construct a gene embedding, we extracted the nucleotide embeddings from positions of pathogenic variants, then we averaged them. All the gene embeddings were stored in a database to be used for the next steps. For loading pre-trained weights, we used the HuggingFace \citep{Wolf2019} interface in Python (\url{https://www.python.org}). For model inference and embedding calculation, we used one Nvidia A100 (40GB) GPU.

\subsection{Case-vs-control analysis}
To assess the impact of pathogenic variants on the gene embeddings, we implemented a case-vs-control approach. For each gene, we trained a logistic regression (with L1 and L2 regularization) using the gene embeddings to classify patients from healthy controls. We used scikit-learn \citep{scikit-learn} to train the model on 75\% of the data and evaluate it with the remaining 25\% resulting in a $F_1$ score for each gene. We compared the gene-specific $F_1$ scores and ranked genes based on this metric (Figure \ref{fig:gene_prioritization_workflow}).

For top candidate gene selection, we used Density-Based Spatial Clustering of Applications with Noise (DBSCAN) \citep{DBSCAN} as an outlier detector. We applied DBSCAN on the calculated $F_1$ scores to find outliers and selected corresponding genes as top candidates.

Finally, to validate the results, we implemented a permutation test. We randomly shuffled the labels (case or control) for N=1000 times. Then we trained a logistic regression on 75\% of the data and calculated a  $F_1$ score on the other 25\%. We counted the number of times that the random $F_1$ score was more than or equal than the observed $F_1$ score. We calculated a p-value as follow (with $\epsilon=0.001$):
\\

$p = \frac{\text{count}(\text{random }F_1 \geq \text{Observed }F_1) + \epsilon}{N + \epsilon}$

\subsection{Case-only analysis}
We also developed a case-only method to prioritize candidate genes if healthy controls are not available. In this approach, for each gene we divided the gene embeddings into mutant (if the patient carried a pathogenic variant) and non-mutant (if the patient was not a carrier). Then we calculated a distance score as the average Euclidean distance between mutant and non-mutant gene embeddings. We utilized these gene-specific distance scores to rank candidate genes (Figure \ref{fig:gene_prioritization_workflow}). 

For top candidate gene selection, similar to the case-vs-control approach, we applied DBSCAN on the distance scores and selected outliers as the top candidate genes. 

To validate the results, we implemented a statistical test as follow : For N=1000 times we generated random reference and alternative embeddings and calculated the distance score. We counted the number of times that the random distance score was more than or equal than the observed distance score. We calculated a p-value as follow (with $\epsilon=0.001$):
\\

$p = \frac{\text{count}(\text{random distance score} \geq \text{Observed distance score}) + \epsilon}{N + \epsilon}$

\subsection{Graph neural network training}

To understand the underlying mechanism of the disease, we designed an explainable approach based on graph neural network (GNN). A summary of the method can be found in figure \ref{fig:pathway_identification_workflow}.
First, we created a protein-protein interaction (PPI) network that indicates interactions between genes carrying pathogenic variants. We used the STRING database \citep{Szklarczyk2023-sb} and included interactions with confidence score $\geq$ 0.6.  

Afterwards, we created individual-specific graphs, which include gene embeddings as node features. We trained a GNN to classify patients’ graphs from controls. GNN architecture consists of two hidden graph convolution layers \citep{Zhang2019} with 16 nodes for message passing and a global sort pooling \citep{Zhang2018} for node feature aggregation. Pooling is essential because the model is trained for graph classification, therefore with pooling we can generate graph representations from node features.
We used AdamW \citep{loshchilov2019decoupled} optimizer with learning rate = 0.001 and weight decay = 0.001 for training. We used batch size = 32 and trained the model for 1000 epochs. We used PyTorch geometric \citep{fey2019fast} for implementing and training the GNN.

\subsection{Subnetwork identification and pathway enrichment analysis}

After training the GNN, we used GNNExplainer \citep{ying2019gnnexplainer} to assign an explainability score to each node, showing how important they are for graph classification . We applied GNNExplainer for all the samples and averaged the explainability scores for each node across samples.

After obtaining the explainability scores, we used the Genetic Algorithm  (GA) \citep{Katoch2020} to identify the “best” subnetwork with maximum fitness, defined as the average of explainability scores of its nodes. GA is a bio-inspired algorithm that mimics evolution by implementing natural selection, chromosomal crossover, and mutation. Previous studies have successfully utilized GA for subnetwork identification \citep{Ulgen2019,Wu2011}. To summarize the GA, we start with a population of random subnetworks, then we select 50\% of subnetworks with probabilities proportional to their fitness scores  (roulette wheel selection). Afterwards, we create new subnetworks by mutating them (adding or removing edges) and crossovering them (connecting two subnetworks, if possible). We started with an initial population of 100 subnetworks and repeated the GA for 10 generations with a mutation rate of 0.5.  At the end, we chose the “most fit” subnetwork at the last generation.

Finally, to gain biological insights into the selected subnetwork, we performed pathway enrichment analysis, a method for identifying biological functions that are over-represented in a group of genes \citep{Chicco2022}. We used the GSEApy package \citep{Fang2022}, which uses Enrichr \citep{Kuleshov2016} for over-representation analysis and the Reactome database \citep{Milacic2023} as reference. We kept significantly enriched pathways with false discovery rate (FDR) $\leq$ 0.05.

\section{Results}

\subsection{Study participants}

As patient cohort (rare disease cases), we used exome data from 120 previously healthy children admitted to PICUs with respiratory failure due to a common viral respiratory infection. Their median age was 78 days, 50 (42\%) were female, and 90 (78\%) were of European ancestry. Respiratory Syncytial Virus (RSV) and Human Rhinovirus (HRV) were the most common detected pathogens, in 67 (56\%) and 31 (26\%) of the cases, respectively.
As controls, we selected 172 healthy individuals from our in-house database of exome-sequenced individuals, representing a random subset of the general population. Since the phenotype we are studying is rare, we assume that the controls are not enriched in individuals with genetic risk factors for infectious disease susceptibility.

\subsection{Variant classification}

In the patient group, 55,300 variants were mapped to coding and splicing regions and were scored with the ACMG/AMP Bayesian classification framework. 48,875 variants had a PoP $\leq$ 0.1 and were considered benign. 5,838 variants had an intermediate PoP (between 0.1 and 0.9), resulting in their classification as variants of unknown significance (VUS). 587 variants (in 508 genes) exceeded the pathogenicity threshold ($\geq$ 0.9) and were considered as damaging.

\subsection{Gene prioritization}

\begin{figure}[t]
  \includegraphics[width=\columnwidth]{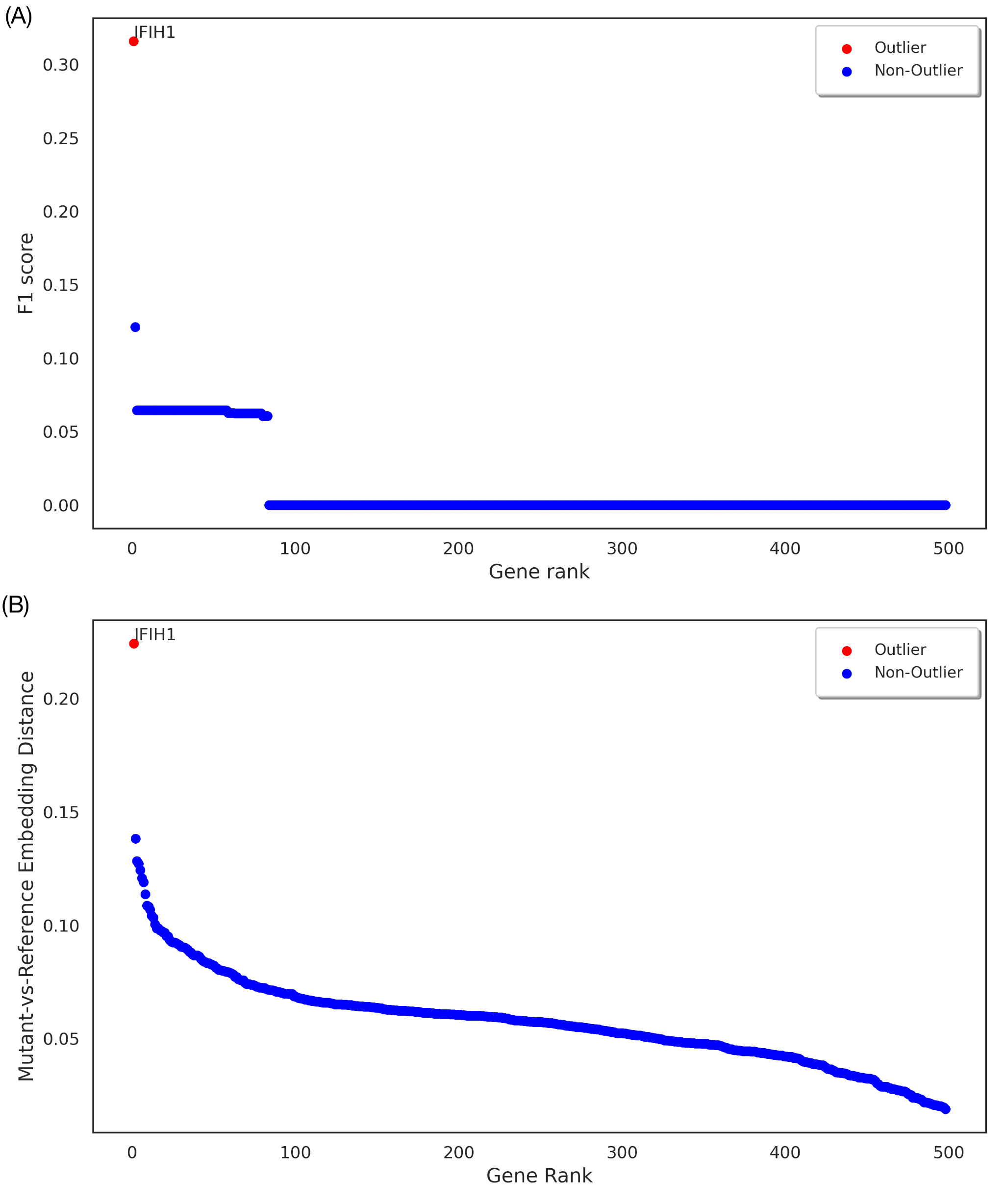}
  \caption{Gene prioritization results. (A) Genes ranked according to their corresponding $F_1$ score calculated based on case-vs-control workflow. (B) Gene ranking based on average distance of mutants and non-mutant embeddings, computed according to the case-only workflow.}
\label{fig:gene_prioritization_results}
\end{figure}

A total of 498 (98\%) candidate genes passed the selection criteria (Methods, Gene embedding calculation). For each candidate gene, we calculated gene embeddings using the pre-trained HyenaDNA for all 292 study participants (120 cases and 172 controls), resulting in gene-specific embeddings in the embedding space. We then ranked candidate genes using two approaches:

1) Case-vs-control: We trained a logistic regression  for each gene and calculated a gene-specific $F_1$ scores. We used these scores to rank the genes and find top candidates by applying DBSCAN for outlier detection. The top candidate gene with the highest $F_1$ score was \textit{IFIH1} (Figure \ref{fig:gene_prioritization_results}.A). We performed a permutation test which resulted in p-value=0.009 (Supplementary figure \ref{fig:F1_permutation}). 

2) Case-only: in this scenario we used the gene-specific distance score (calculated based on the average Euclidean distance of mutant and non-mutant embeddings) for gene prioritization and top-candidate selection. \textit{IFIH1} ranked first  and was selected as an outlier using the DBSCAN method (Figure \ref{fig:gene_prioritization_results}.B) and was significantly different from the expected distribution (p-value=$10^{-6}$, Supplementary figure \ref{fig:distance_permutation}).

\subsection{PPI construction and graph neural network training}

We constructed a high-quality PPI based on the interactions between the protein products of all candidate genes, resulting in a PPI with 138 nodes and 176 edges. For each participant, we initialized the same PPI structure, but used their personalized gene-embeddings as node features, resulting in 292 (120 cases and 172 controls) unique graphs.
We used these graphs to train a GNN for classifying cases from controls. GNN structure consisted of 2 graph convolution layers with 16 nodes, and global sort pooling to generate graph representations from node features. We trained the GNN for 1000 epochs.

\subsection{Subnetwork identification and pathway enrichment analysis}

\begin{figure}[t]
  \includegraphics[width=\columnwidth]{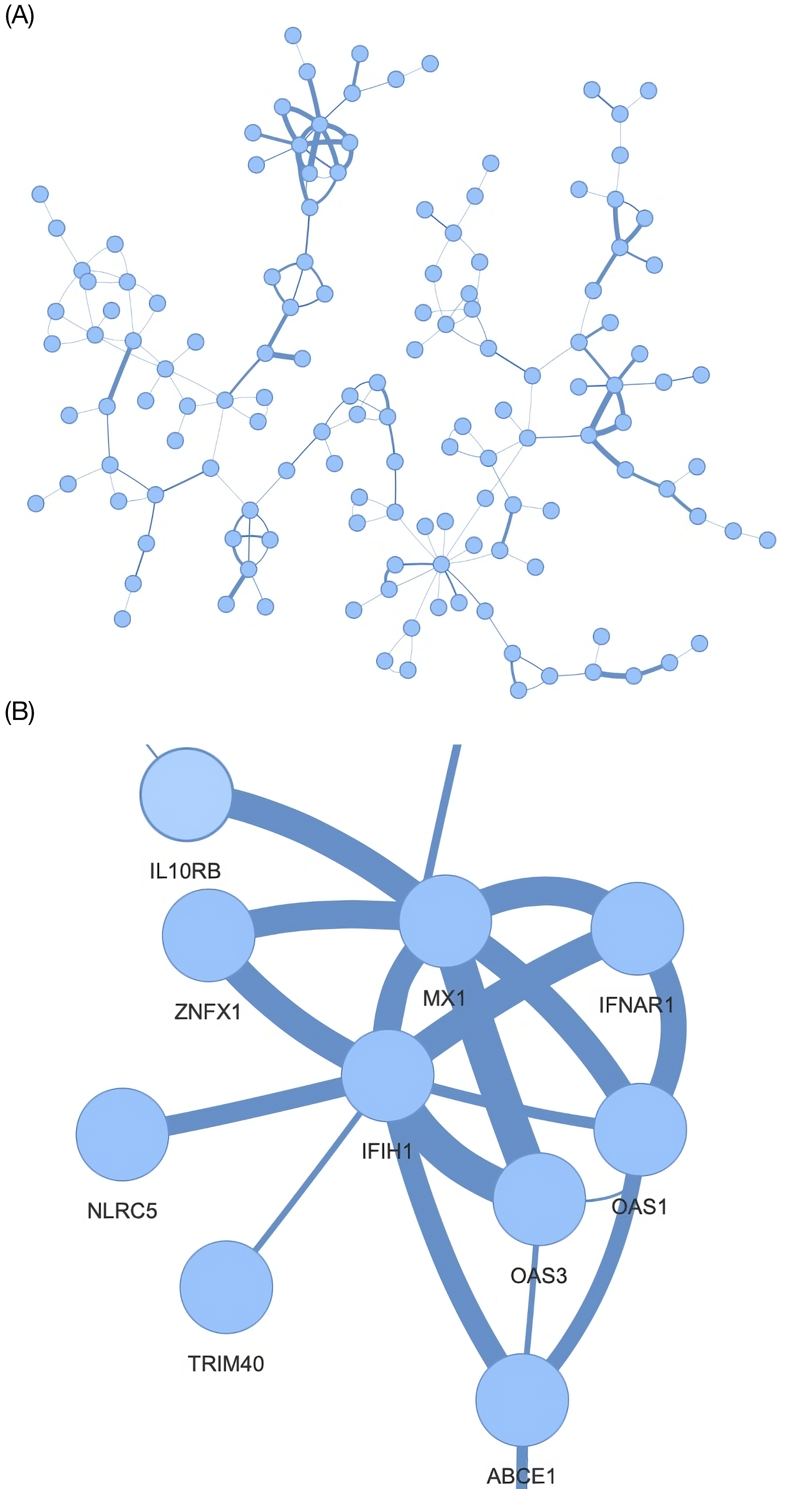}
  \caption{Subnetwork identification results. (A) PPI of candidate genes scored using GNNExplainer. The thickness of edges reflects the importance of nodes connected to it. (B) Selected subnetwork with maximum fitness, defined as the average nodes’ importance scores. This subnetwork is identified via the Genetic Algorithm.}
\label{fig:subnetwork_identification_results}
\end{figure}

\begin{figure}[t]
  \includegraphics[width=\columnwidth]{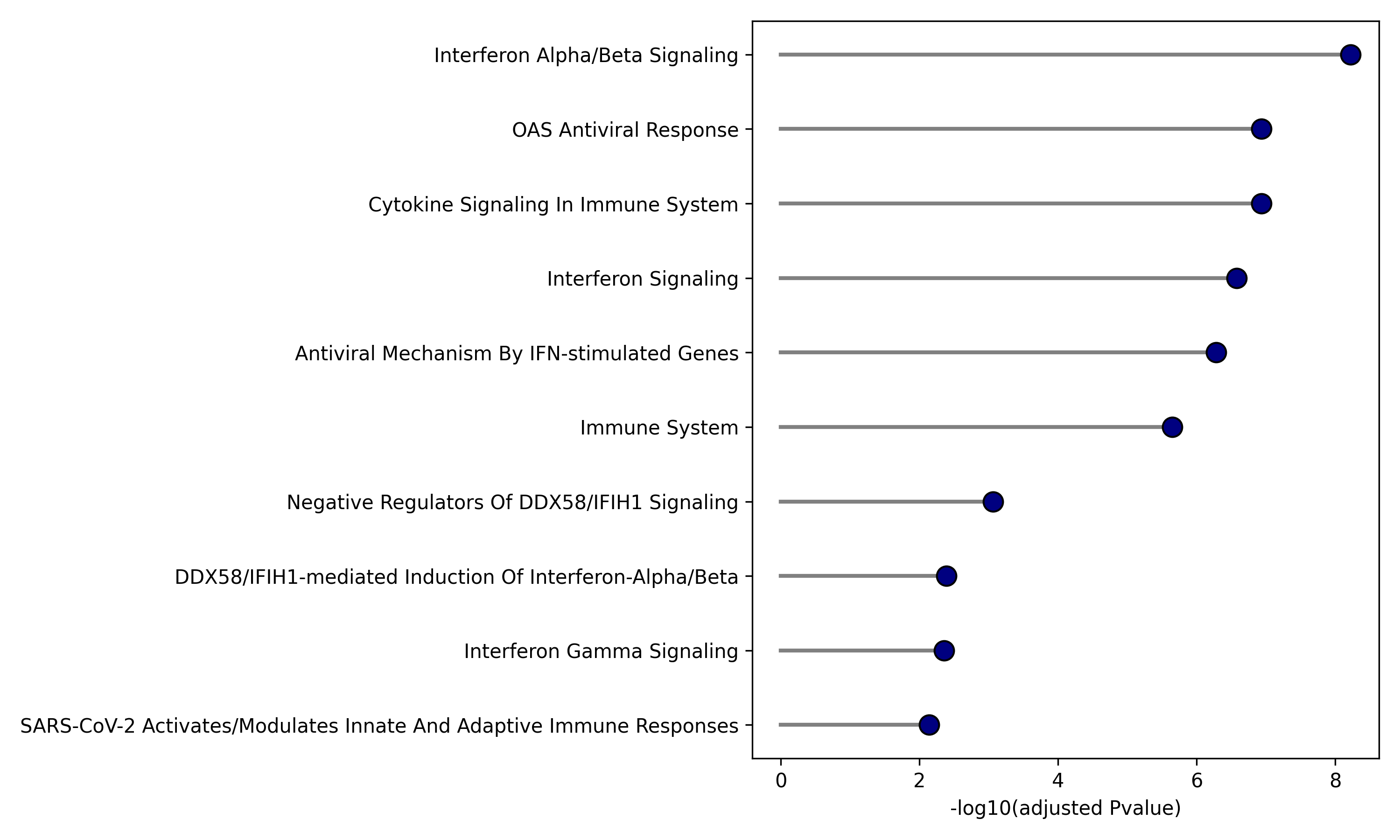}
  \caption{Top 10 significantly enriched pathways using the Reactome database. Genes in the selected subnetwork were used as input.}
\label{fig:pathway_enrichment_results}
\end{figure}

After training the GNN, we used GNNExplainer to assign an explainability score to each node, showing how important they are for graph classification. We applied GNNExplainer for all the samples and  averaged the explainability scores for each node across samples. Figure \ref{fig:subnetwork_identification_results}.A shows the PPI with explainability scores reflected on the edges’ widths.
After obtaining the explainability scores, we used the Genetic Algorithm to identify the “best” subnetwork with maximum fitness. The fitness of a subnetwork was defined as the average of explainability scores of its nodes . This resulted in a subnetwork with 10 genes including \textit{IFIH1}, \textit{OAS1}, \textit{OAS3}, \textit{MX1}, \textit{IFNAR1}, \textit{IL10RB}, \textit{ZNFX1}, \textit{NLRC5}, \textit{TRIM40}, and \textit{ABCE1} (Figure \ref{fig:subnetwork_identification_results}.B). 
Finally, we performed pathway enrichment analysis using the Reactome database as reference and kept significantly enriched pathways with FDR $\leq$ 0.05. Top 10 resulting pathways are shown in figure  \ref{fig:pathway_enrichment_results}. 

\section{Discussion}

In this study we aim to harness the potential of DNA foundation models to translate the intricate 'language' of DNA into meaningful and actionable information. We propose a framework to utilize DNA-LMs for gene prioritization and pathway identification in rare disease studies. Based on the hypothesis that variants with strong deleterious effects alter the gene embeddings significantly in the embedding space, we demonstrate that it is possible to prioritize disease-associated genes/pathways in a cohort of 120 children requiring intensive care support because of a severe illness caused by a respiratory virus.

For gene prioritization, we propose two approaches to analyze the gene embeddings (Figure \ref{fig:gene_prioritization_workflow}): case-vs-control and case-only. The case-only approach is particularly promising for rare disease research, where finding a well-matched control group is often challenging. The ability of the method to differentiate between mutant and non-mutant gene embeddings within the same patient cohort is a novel and practical solution to this long-standing issue. By applying the gene prioritization workflow, we successfully re-identified \textit{IFIH1} - which encodes an RIG-I-like receptor involved in the sensing of viral RNA \citep{Rehwinkel2020} - as the top candidate gene in our patient cohort.

For pathway identification, we propose an integrative method, combining DNA-LM with interpretable GNN and Genetic Algorithm (Figure \ref{fig:pathway_identification_workflow}). This approach takes into account various information such as PPI, number of variant carriers, and context-specific impact of variants on gene sequences. By applying this method, we were able to identify potentially relevant genes (\textit{IFIH1}, \textit{OAS1}, \textit{OAS3}, \textit{MX1}, \textit{IFNAR1}, \textit{IL10RB}, \textit{ZNFX1}, \textit{NLRC5}, \textit{TRIM40}, and \textit{ABCE1}) that can explain the disease pathogenesis. 

All the identified genes are coding for molecules that play an important role in antiviral defense. \textit{IFIH1} encodes MDA5, which is a cytoplasmic viral RNA sensor that recognizes single- or double-strand RNA to launch a type 1 interferon response \citep{Rehwinkel2020}. \textit{OAS1} and \textit{OAS3} encode enzymes that activate host RNase L to degrade viral RNA \citep{Hornung2014}. \textit{ABCE1} encodes a protein that is involved in the regulation of OAS/RNase L pathway \citep{Martinand1998}. \textit{MX1} encodes a guanosine-triphosphate-metabolizing protein that antagonizes the replication process of viruses \citep{Haller2019}. \textit{IFNAR1} and \textit{IL10RB} encode cytokine receptors that mediate the antiviral immunity \citep{Zanin2021,Moore2001}. \textit{ZNFX1} encodes a protein that binds to viral RNA and interacts with mitochondrial antiviral signaling (MAVS) protein, promoting the expression of interferon-stimulated genes \citep{Vavassori2021}. \textit{NLRC5} and \textit{TRIM40} encode regulators of antiviral signaling pathways \citep{Kuenzel2010,Zhao2017}. Deficiencies in some of these genes have been previously studied and shown to impair immunity against specific human viruses \citep{Lamborn2017,Asgari2017,Chen2021-yz,Abolhassani2022-cg,Korol2023-kc,Saadat2023,Lee2023-tk}.  

In this study we focused on DNA-LMs, although protein language models (pLMs) such as ESM-1b \citep{Brandes2023} have demonstrated state-of-the-art performance in scoring missense variants. The reason we used a DNA-LM instead of pLM is that DNA-LMs can model various variant types (e.g., splicing, stop-gained, etc.) while pLMs focus only on missense variants. Moreover, by using DNA-LMs, our method can be extended to other variant types such as those mapping to introns, branchpoint motives, or untranslated regions (UTRs).

While our method shows promise, there are inherent challenges and limitations. Our proposed workflow identifies genes with significant changes in their embeddings, yet a careful analysis is required to quantify the minimum embedding distortion to be detectable by the model. Moreover, the interpretation of gene embeddings requires careful consideration, since not all genetic variations captured in the embeddings might be clinically relevant.  

The potential for integrating DNA-LMs with other techniques, such as multi-omics, could further enhance our understanding of genetic diseases. This has significant implications for the identification of disease-causing genes/pathways, potentially leading to more targeted and effective treatments in personalized medicine. The demonstration that DNA-LMs can accurately identify genes and pathways involved in rare diseases paves the way for further research and application of artificial intelligence in various genomics research domains.

\section*{Code Availability}
The code for this study is available \href{https://github.com/AliSaadatV/Causal-Gene-Pathway-Finder}{here}. 

\bibliography{acl_latex}

\appendix
\onecolumn
\section{Appendix}
\label{sec:appendix}

\newcounter{suppfigure}
\renewcommand{\thefigure}{S\arabic{suppfigure}} 
\setcounter{suppfigure}{1}
\renewcommand{\figurename}{Supplementary Figure}

We used ACMG/AMP guidelines \citep{Richards2015} to classify the variant into putative pathogenicity groups, as described in our previous works \citep{Saadat2023,saadat-fellay-2024-dna,saadat2024finetuning,saadat2024finetuningesm2proteinlanguage}. In summary, we gather all the available evidences for a variant. Table \ref{tab:tab1} summarizes all the ACMG/AMP criteria that we used.

\begin{table}[ht]
\centering
\includegraphics[width=\textwidth]{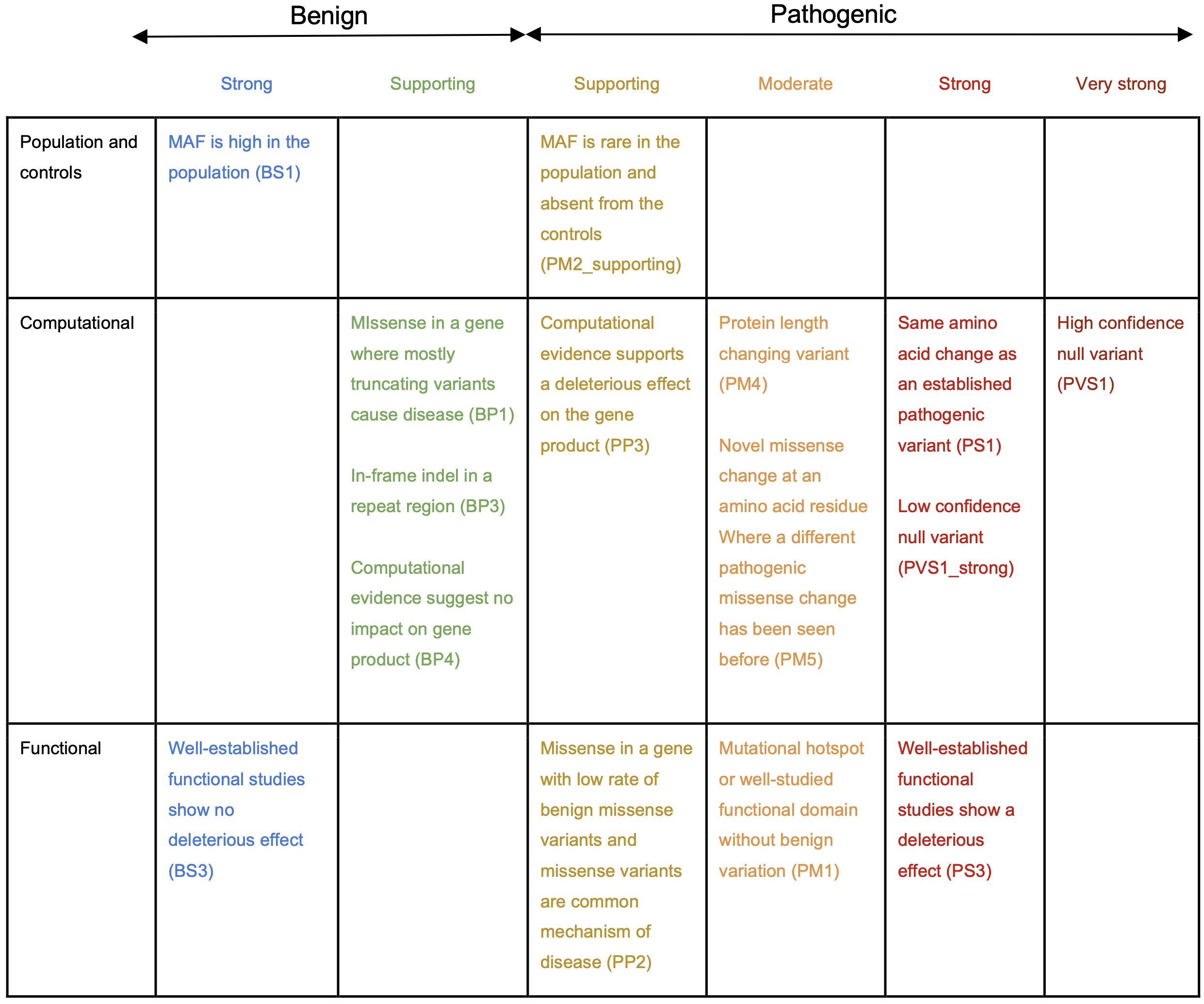}
\caption{the summary of ACMG/AMP criteria used for variant classification. MAF: minor allele frequency}
\label{tab:tab1}
\end{table}

To calculate the probability of pathogenicity (PoP), we use the Bayesian framework developed by \citet{Tavtigian2018}. For a given variant, the PoP is calculated as follow:

\begin{center}
$P_x = \text{number of pathogenic criteria applied at the level of } x$ \\
$x \in \{\text{Very strong}, \text{Strong}, \text{Moderate}, \text{Supporting}\}$ \\
[1em]

$B_y = \text{number of benign criteria applied at the level of } y$ \\
$y \in \{\text{Strong}, \text{Supporting}\}$ \\
[1em]

$\text{odds of pathogenicity (OP)} = 350^{(\frac{P_{\text{Very strong}}}{1} + \frac{P_{\text{Strong}}}{2} + \frac{P_{\text{Moderate}}}{4} + \frac{P_{\text{Supporting}}}{8} - \frac{B_{\text{Strong}}}{2} - \frac{B_{\text{Supporting}}}{8})}$ \\
[1em]

$\text{probability of pathogenicity (PoP)} = \frac{OP \times 0.1}{((OP - 1) \times 0.1 + 1)}$
\end{center}

\begin{figure}[t]
  \includegraphics[width=\columnwidth]{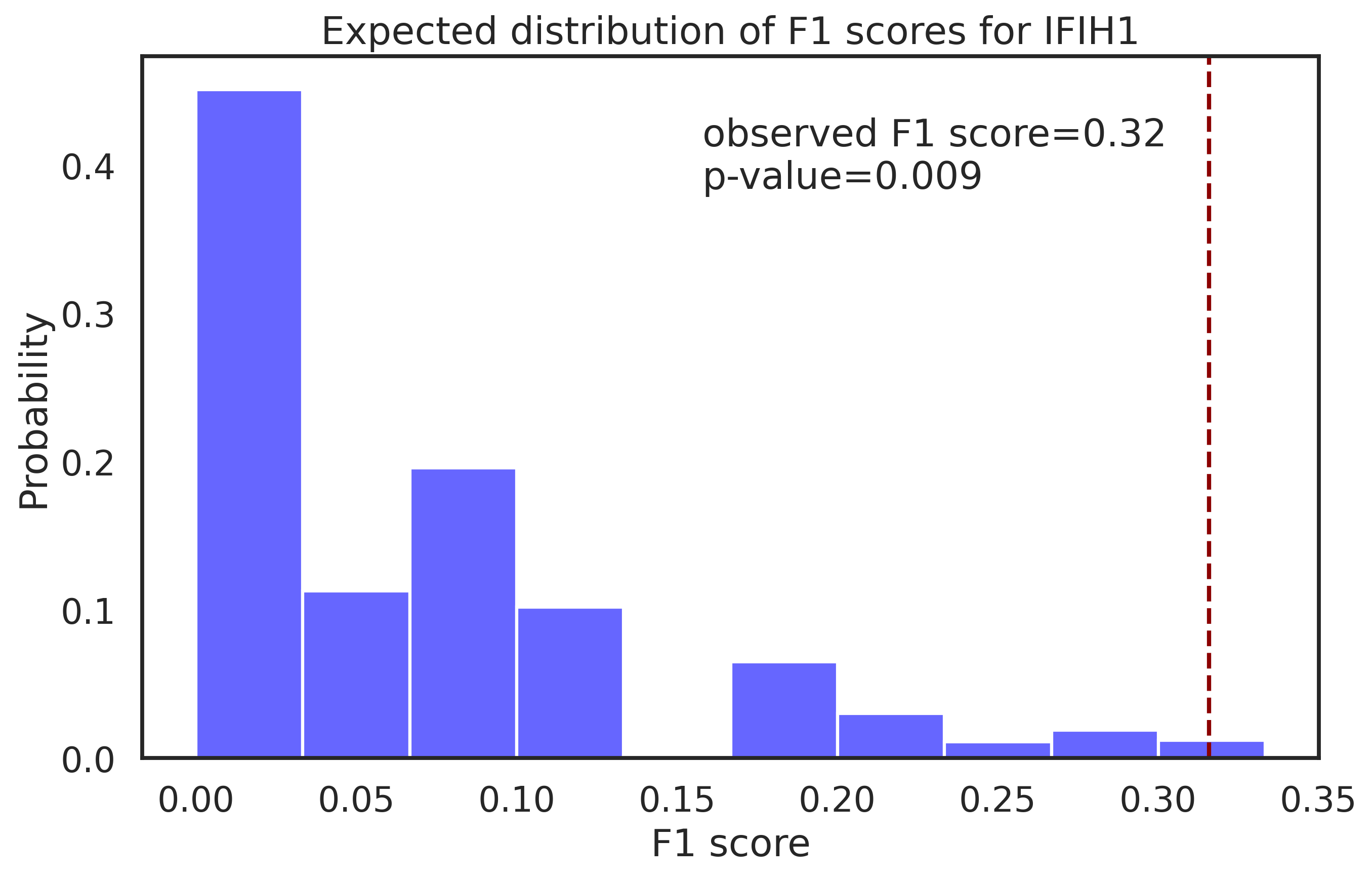}
  \caption{Permutation test results for case-vs-control approach. Expected distribution of $F_1$ scores for \textit{IFIH1} is shown in blue. The red line indicates the observed $F_1$ score.}
\label{fig:F1_permutation}
\refstepcounter{suppfigure}
\end{figure}

\begin{figure}[t]
  \includegraphics[width=\columnwidth]{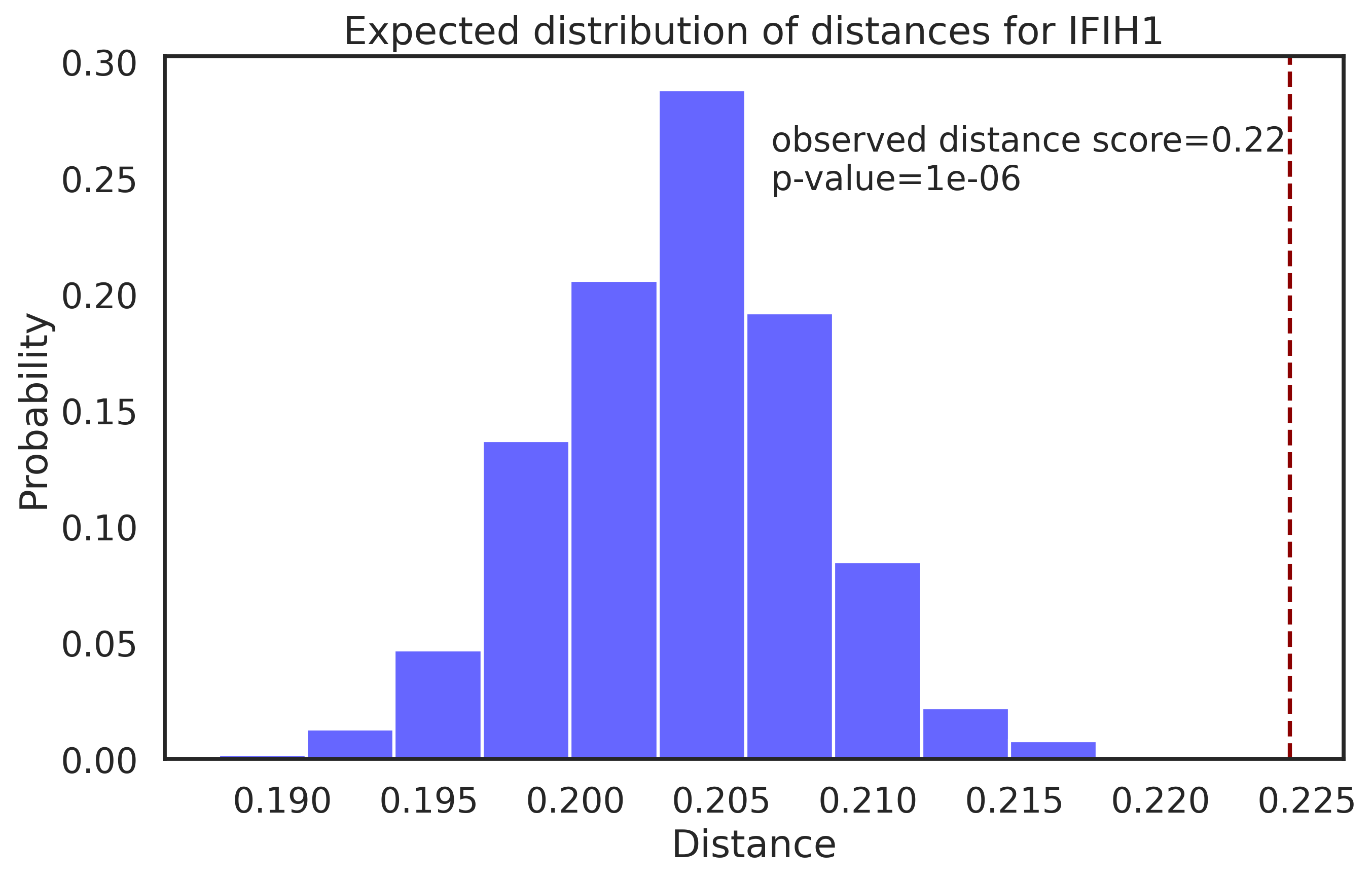}
  \caption{Statistical test results for the case-only approach. Expected distribution of distance scores for \textit{IFIH1} is shown in blue. The red line indicates the observed distance score.}
\label{fig:distance_permutation}
\refstepcounter{suppfigure}
\end{figure}

\end{document}